\documentclass[conference]{IEEEtran}
\IEEEoverridecommandlockouts
\usepackage{cite}
\usepackage{amsmath,amssymb,amsfonts}
\usepackage{algorithmic}
\usepackage{graphicx}
\usepackage{textcomp}
\usepackage{xcolor}
\def\BibTeX{{\rm B\kern-.05em{\sc i\kern-.025em b}\kern-.08em
    T\kern-.1667em\lower.7ex\hbox{E}\kern-.125emX}}
\begin{document}

\title{A Dive into WhatsApp's End-to-End Encryption}

\author{\IEEEauthorblockN{Pinaki Prasad Guha Neogi}
\IEEEauthorblockA{\textit{Dept. of Computer Science, School of ENCS} \\
\textit{Washington State University}\\
Washington, USA \\
p.guhaneogi@wsu.edu}
}

\maketitle

\begin{abstract}
We live in a generation where the world around us is witnessing technological revolutions every single day, and as a result of this, everything around us is getting digitized with the touch of technology. In order to keep up the pace of this technological revolution and help reaching this progress its zenith, one of the most important aspects that needs to be taken care of is security. One of the biggest boons of technology in the recent times has been the invention of smartphones. As smartphones started becoming more popular, affordable and easily accessible, hundreds of free messaging applications were launched, but WhatsApp emerged as the ultimate winner in the race. This paper describes one of the most important and popular features of WhatsApp, the End-to-End (E2E) encryption system, which sets it apart from most other messaging applications and is one of the reasons which helped it become so popular.
\end{abstract}


\section{Introduction}
Though the world around us has always been witnessing changes as a consequence of the drastic advancement in the field of science and technology, these days it’s been like never before. Nowadays, it has become almost impossible to escape the presence of technology in our day-to-day lives. Smartphone is one such thing that has revolutionized our lives, and ever since its popularity kept increasing, hundreds of messaging applications have been launched over the past couple of years. Out of these, WhatsApp, which is a free messaging application owned by Meta Inc. (formerly known as Facebook Inc.) became the most popular one with more than 5 billion users in over 180 nations \cite{b1} .  

WhatsApp was created in the year 2009 by Jan Koum and Brian Acton with the intention of making communication as well as media sharing faster and more convenient \cite{b2}. WhatsApp requires internet connection for its operations, and it enables the users to stay in touch with family members and friends on their contact list only, i.e., in order to connect with someone over WhatsApp his/her contact number needs to be saved in the user’s phone (unlike Facebook and most other social messaging applications, where one can connect with unknown people as well). In addition to offering all the perks of free social media messaging applications, it also helps keeping the circle private, thus making it even more popular. Besides text messages, it also offers features like creating groups, calling, video calling, sharing multimedia files like images, audios, documents and videos etc. \cite{b3}. It also helps users understand if a particular message has been sent, delivered and read by the recipient. Also, a user can block a particular person on his contact list if he wishes. 

As more and more people started opting for WhatsApp as their mode of communication, ensuring security of the communication channels became very important because a fair amount of privacy is expected by the users whenever they use a communication medium, and no one wants to use an application with questionable privacy for private or business communications. With the aim of fulfilling this expectation, in the year 2014, WhatsApp came forward with the concept of End-to-End (E2E) Encryption. This ensures the security of the data shared between the two parties, making it hard to crack and free from eavesdropping. Fig. 1 shows the End-to-End encryption policy of WhatsApp, as it appears on the app. This makes communication more reliable because as per the E2E encryption principles, the data is safe in transit, and no one other than the intended recipient (not even WhatsApp itself) can decrypt and read the data. However, in spite of ensuring security, privacy and integrity, it knocks out government surveillance for the sake of the country’s safety against threats like terrorism. 

\begin{figure}[htbp]
\centerline{\includegraphics[width=56mm]{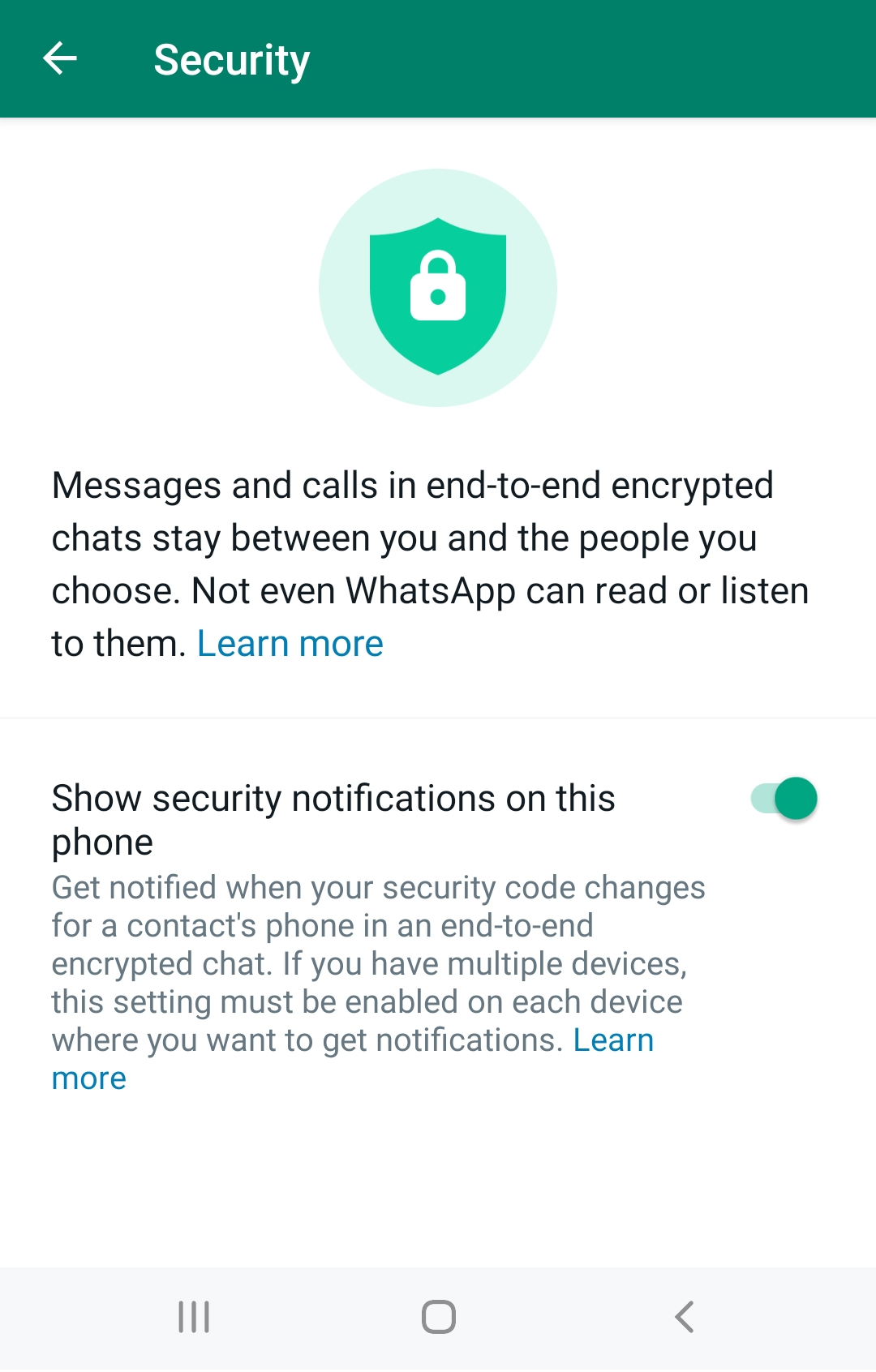}}
\caption{WhatsApp's End-to-End Encryption Policy.}
\label{fig1}
\end{figure}

\section{Features of a Secure messaging system}

There are few minimum features that need to be ensured in order to call a messaging system a secure one. Suppose we have a messaging system where Adam and Bud want to exchange messages with each other, and there is a third person, Mark, who attempts mid-in-the-middle attack and wishes to listen to their conversation and possibly forge the communication by sending messages to Bud acting as Adam, and to Adam acting as Bud. These characters have been chosen based on the roles of Alice, Bob and Mallory – commonly used characters in cryptography literature. Amongst many others, we expect at least the following main features to be ensured in a secure messaging system:

\begin{itemize}
\item \textbf{Confidentiality:} Mark should not know what messages Bud and Adam exchange with each other. This could be achieved with the help of some cryptographic techniques (encryption/decryption).

\item \textbf{Integrity:} it prevents messages from being modified. When Adam sends some message to Bud, Bud can check whether the message was subjected to any modification (by Mark) on its way. Digital signature can be employed in order to ensure data integrity.

\item \textbf{Authenticity:} authenticity refers to the way of ensuring upon receiving a message from Adam that it was actually sent by Adam, and not Mark.  This could be achieved with the help of digital signatures and MACs.

\end{itemize}
\vspace{2mm}
Given the requirement for secure communication, the next most important thing is to decide which type of cryptography should be used, which can be broadly classified into two categories – symmetric key cryptography and asymmetric (public) key cryptography. In symmetric key cryptography, the same key is used for encrypting and decrypting a particular message and hence it requires a secure way of exchanging this secret key. Whereas, in asymmetric key cryptography, anyone can encrypt a message using the public key, and only the intended recipient can decode it with the private key. Asymmetric cryptography is convenient because it doesn’t involve sharing the secret key as long as the users know each other’s public key. The only concern is to make sure that they are using the correct public key e.g., if Adam is sending a message to Bud, he needs to be assured that he is using Bud’s public key and not Mark’s. Even if Mark learns about Adam and Bud’s public keys, it’s okay, because they are public. Fig. 2 shows the way to verify the E2E encryption security code between two parties (say, Adam and Bud). It can be verified either by scanning the QR code or by matching the security codes manually. 

\begin{figure}[htbp]
\centerline{\includegraphics[width=56mm]{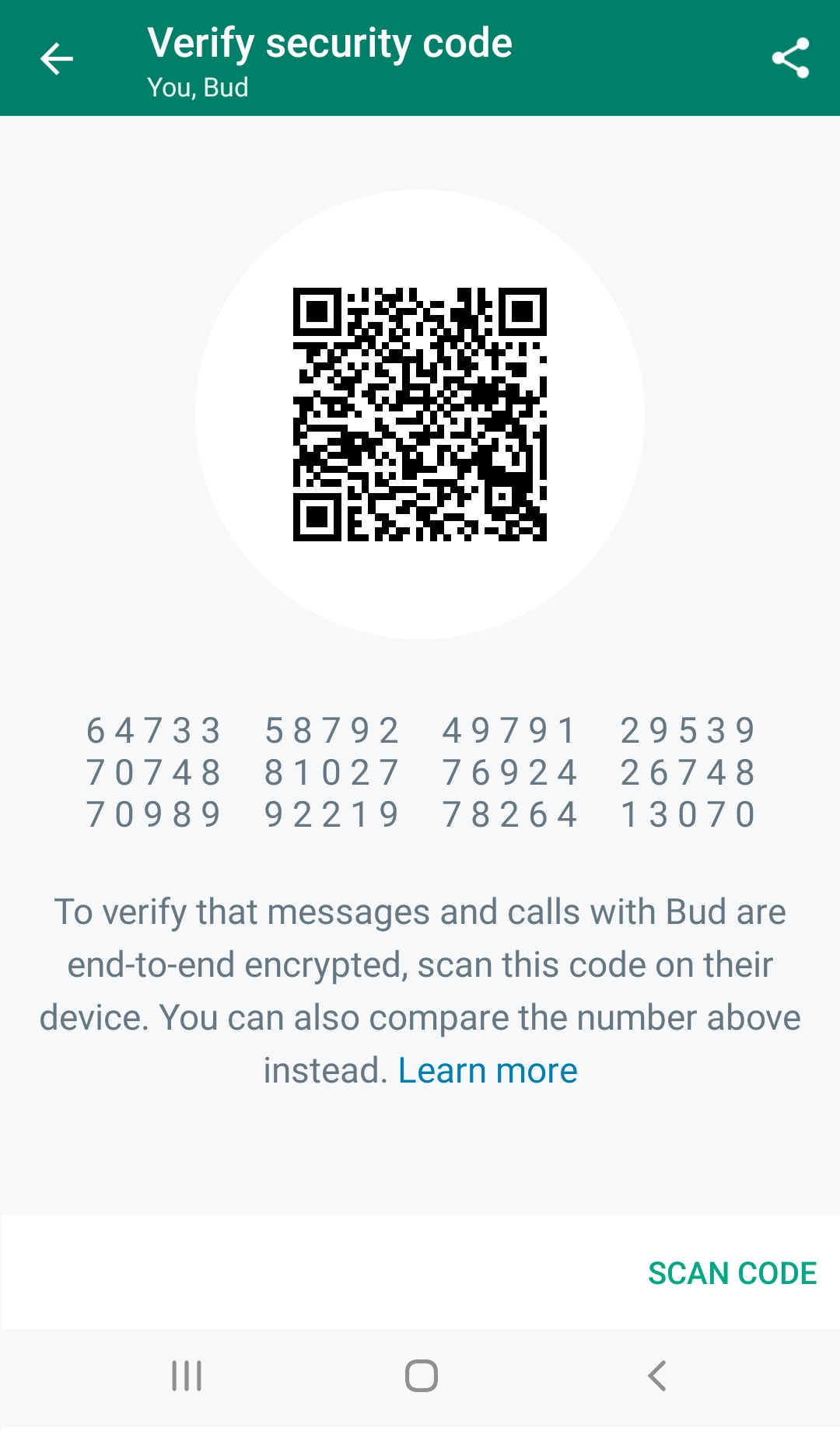}}
\caption{Security Code Verification.}
\label{fig2}
\end{figure}

\section{WhatsApp’s E2E Encryption protocol}
WhatsApp employs the open source Signal Protocol for building its End-to-End (E2E) encryption system. This is the same protocol that is being used by the signal messaging application and was initially developed by the Open Whisper Systems. The signal protocol makes use of primitives such as prekeys, Curve25519, Diffie Hellman Key Agreement, Extended Triple Diffie Hellman Key Agreement (X3DH), AES, HMAC\_SHA256 and Double Ratchet Algorithm. Before putting everything together and explain how the protocol functions, let’s have a look into each of these primitives one by one: 

\subsection{Curve25519}\label{AA}
In the field of cryptography, Curve25519 basically refers to either a cryptographic system based on Edwards curve 25519 or the elliptic curve relation itself which, with a key size of 256 bits, offers 128 bits of security and has been primarily designed for employing it with the elliptic curve Diffie–Hellman (ECDH) key agreement protocol. It is considered to be one of the fastest Elliptic curve cryptographies. The original paper describes Curve25519 as a Diffie-Hellman (DH) function, and the author Daniel J. Bernstein suggested the name Curve25519 for the underlying elliptic curve and X25519 for the corresponding Diffie-Hellman function \cite{b4}.

\subsection{Prekeys}

Prekeys are basically Curve25519 key pairs which are used for message authentication and for secure exchange of messages.  There are basically three types of prekeys:
\begin{enumerate}
\item \textbf{Identity keys (IK):} these are also known as long-term identity keys because they are generated at the time of installing the application into a device. These keys are permanent until the device is changes or the app reinstalled. 

\item \textbf{Signed Prekeys (SPK):} unlike identity keys, signed prekeys are not permanent, but they are not disposable either. These keys have a specific lifetime and hence, need to be updated periodically. 

\item \textbf{One-Time Prekeys (OPK):} these keys are disposable and they get deleted from the server each time some user (who wants to send some messages to him) requests a public key bundle from the server. As the name suggests, each of these keys is used for a for a single Extended Triple Diffie Hellman (X3DH) protocol run. 
\end{enumerate}

\subsection{Diffie Hellman Key Agreement}
Diffie Hellman Key Agreement, sometimes also known as Diffie-Hellman-Merkle Key Agreement is a key exchange mechanism for implementing asymmetric key cryptography, where both the parties involved in the communication agree on a set of protocol parameters (a prime base \(B\) and a prime modulus \(G\)) to be used for communication over a public channel. These parameters are public, and are known to everyone. Let’s understand the working principle of the Diffie-Hellman key exchange mechanism with the help of an example. 

Say for example, Adam and Bud want to exchange some messages with each other and they want to use the Diffie-Hellman key exchange mechanism for generating the same shared secret. The steps involved are as follows:
\begin{itemize}
\item Suppose both Bud and Adam agree to use base \(B\) and modulus \(G\)
\item Adam selects a random value \(x\) and keeps it secret (Adam’s private key)
\item Bud selects a random value \(y\) and keeps it secret (Bud’s private key) 
\item Adam computes \(X = B^x \: mod \: G\) and shares it with Bud (Adam’s public key)
\item Bud computes \(Y = B^y \: mod \: G\) and shares it with Adam (Bud’s public key)
\item Upon receiving Bud’s public key, Adam computes the shared secret \(S_x = Y^x \: mod \: G\)
\item Upon receiving Adam’s public key, Bud computes the shared secret \(S_y = X^y \: mod \: G\)
\end{itemize}
\vspace{2mm}
The shared secret computed by both Adam and Bud happens to be equal, both being \(B^{xy} \: mod \: G\). Even if Mark attempts an attack, it is computationally infeasible for him to determine the actual values of the private keys (\(x\) and \(y\)) by just knowing the public keys (\(X\) and \(Y\)).

\subsection{Extended Triple Diffie Hellman Key Agreement (X3DH)}

X3DH is an extension of the Diffie-Hellman protocol, explained in the previous section, with the aim of overcoming its limitations for the asynchronous setting. Suppose, Adam wishes to establish a secure mode of communication with Bud by exchanging the keys. The algorithm explains above would work perfectly as long as both Adam and Bud are online. But what if Bud is offline? There is no other way than to wait until Bud is back. \cite{b8} Also, there is no way for Adam and Bud to validate if they are talking to each other and using each other’s keys (and not using Mark’s keys, thinking they are Adam’s or Bud’s). X3DH protocol overcomes these limitations. 

Assuming Bud being the receiver, and Adam the sender, the working principle of X3DH can be explained in the following three major phases:
\begin{enumerate}
\item Publishing keys.
\item Sending the initial message.
\item Receiving the initial message.
\end{enumerate}
\vspace{2mm}
\begin{enumerate}
\item \textbf{Publishing keys:} Bud, the receiver, publishes a set of Curve25519 public keys to the server before the communication begins. This includes the public keys of the following key pairs:
\vspace{2mm}
\begin{itemize}
\item Bud’s identity key (\(IK_b\))
\item Bud’s signed prekey (\(SPK_b\))
\item Bud’s prekey signature, which can be represented as \(Sign (Encode (SPK_b), IK_b)\)
\item Set of one-time prekeys which can be denoted as (\(OTPK_{b1}, \: OTPK_{b2}, \: OTPK_{b3}, …\))
\end{itemize}
\vspace{1mm}
The identity key \((IK_b)\) is uploaded to the server only once (or not until the app is reinstalled or the device is changed). The signed prekey \((SPK_b)\) and the prekey signature \(Sign (Encode (SPK_b), IK_b)\) i.e., \(SPK_b\) signed with the identity key \((IK_b)\) need to be uploaded at a regular interval – once a week or once a month. And new one-time prekeys are uploaded whenever server’s collection of the users’ one-time prekeys is getting low. As new keys are uploaded, older ones get deleted (but they are stored for some time so as to decode any message encrypted with the older keys).
\vspace{2mm}
\item \textbf{Sending the initial message:} In order to perform an extended triple Diffie-Hellman key agreement with Bud, Adam fetches a “prekey bundle” from the server containing the public keys of the following key pairs:
\vspace{2mm}
\begin{itemize}
\item Bud’s identity key \((IK_b)\)
\item Bud’s signed prekey \((SPK_b)\)
\item Bud’s prekey signature, which can be represented as \(Sign (Encode (SPK_b), IK_b)\)
\item (Optional) Any of Bud’s one-time prekeys \((OTPK_b)\)
\end{itemize}
\vspace{1mm}
The one-time prekey is optional for sending the initial message. The bundle will contain a one-time prekey if there exists one in the server, otherwise it will not contain a one-time prekey.
\vspace{1mm}
Upon receiving the prekey bundle, Adam first verifies the prekey signature \(Sign (Encode (SPK_b), IK_b)\). If the verification fails, the protocol is aborted, otherwise Adam generates a new pair of ephemeral keys (let’s represent the public key of the pair as \(EK_a)\).
\vspace{1mm}
Now, let’s consider two possible cases – case 1: when the prekey bundle contains a one-time prekey; case 2: when there is no one-time prekey in the prekey bundle. Depending upon the scenario, i.e., to which case it falls, Adam computes the secret key \((SK)\) as follows:
\vspace{2mm}
\subsubsection{Case 1}
\begin{equation}
DH_1 = DH (IK_a, \: SPK_b)\label{eq1}
\end{equation}
\begin{equation}
DH_2 = DH (EK_a, \: IK_b)\label{eq2}
\end{equation}
\begin{equation}
DH_3 = DH (EK_a, \: SPK_b)\label{eq3}
\end{equation}
\begin{equation}
SK = KDF (DH_1\: || \: DH_2 \:|| \:DH_3)\label{eq4}
\end{equation}

-where,
\begin{itemize}
\item The function \(DH (K1, K2)\) returns the shared secret computed by the Elliptic Curve Diffie-Hellman function, using public keys \(K1\) and \(K2\). 
\item The function \(KDF(x)\) returns a 32-byte output by feeding \(x\) to the HMAC-based Extract-and-Expand Key Derivation Function (HKDF) algorithm \cite{b5}.
\item \(A \:||\: B\) represents the byte concatenation of the sequences \(A\) and \(B\).
\end{itemize}

\vspace{2mm}
\subsubsection{Case 2}
\begin{equation}
DH_1 = DH (IK_a, \: SPK_b)\label{eq5}
\end{equation}
\begin{equation}
DH_2 = DH (EK_a, \: IK_b)\label{eq6}
\end{equation}
\begin{equation}
DH_3 = DH (EK_a, \: SPK_b)\label{eq7}
\end{equation}
\begin{equation}
DH_4 = DH (EK_a, \: OTPK_b)\label{eq8}
\end{equation}
\begin{equation}
SK = KDF (DH_1\: || \: DH_2 \:|| \:DH_3\:|| \:DH_4)\label{eq9}
\end{equation}

\vspace{2mm}
After the secret key (SK) has been computed, Adam deletes the DH outputs along with his ephemeral private key. And then computes an “associated data” byte \((AD)\), comprising the identity keys of both Adam and Bud:
\begin{equation}
AD = Encode (IK_a) || Encode (IK_b)\label{eq8}
\end{equation}

Following this, Adam sends Bud an initial message comprising:
\begin{itemize}
\item Adam’s identity public key \((IK_a)\)
\item Adam’s ephemeral public key \((EK_a)\)
\item Identifiers indicating which of the Bud’s prekeys have been used by Adam
\item An initial ciphertext, obtained by passing the associated data \((AD)\) and an encryption key (the secret key \(SK\) computed above, or the output from some Pseudo-Random Functions keyed by \(SK\)) through some AEAD encryption scheme \cite{b6}. This ciphertext basically served a dual role – besides being Adam’s initial X3DH message, it also acts as the initial message within a number of post-X3DH protocols.  
\end{itemize}

\vspace{2mm}
\item \textbf{Receiving the initial message:} Upon receiving the initial message sent by Adam, Bud retrieves Adam’s identity public key \((IK_a)\) and ephemeral public key \((EK_a)\) from the message. At the same time, he also loads his own identity private key, and the private counterparts of the signed prekey and the one-time prekey used by Adam. Using these keys that Bud loads and retrieves, he himself repeats the DH and KDF computations described in the previous section and obtain the secret key (SK).
Bud then constructs the associated data byte (AD) using the identity keys \(IK_a\) and \(IK_b\), and following the same approach used by Adam as described in the previous section. 

And finally, after obtaining both the SK and the AD values, Bud tries to decode the initial ciphertext sent by Adam. If the decoding fails, Bud deletes the secret key (\(SK\)) and terminates the protocol. Otherwise, if Bud could successfully decode the initial ciphertext, the protocol is settled for him.  In that case, Bud deletes the private key corresponding to the one-time prekey used by Adam for sending the initial message, but he may continue using the secret key (\(SK\)) for post-X3DH computations and processing.
\end{enumerate}

\subsection{HMAC\_SHA256}
HMAC\_SHA256, as the name suggests, is a Hash-based Message Authentication Code (HMAC) which is designed based on the principle of SHA-256 hash function. This method encrypts the message data with the secret key \((SK)\), hashes the output by passing it through the hash function, then again combines the secret key \((SK)\) with this resulting hash value, and hashes it for a second time. This yields a 256-bit output hash. This technique is actually used to determine if a message has been subjected to tampering while being transferred from a sender to the receiver.

\subsection{Double Ratchet}
The term ratchet refers to a mechanical device which can move in only one direction (either rotates from left to right or from right to left, but not in both directions). Double ratchet, as the name indicates is an amalgamation of two different ratchets – the symmetric-key ratchet and the Diffie-Hellman ratchet. 

\begin{figure}[htbp]
\centerline{\includegraphics[width=70mm]{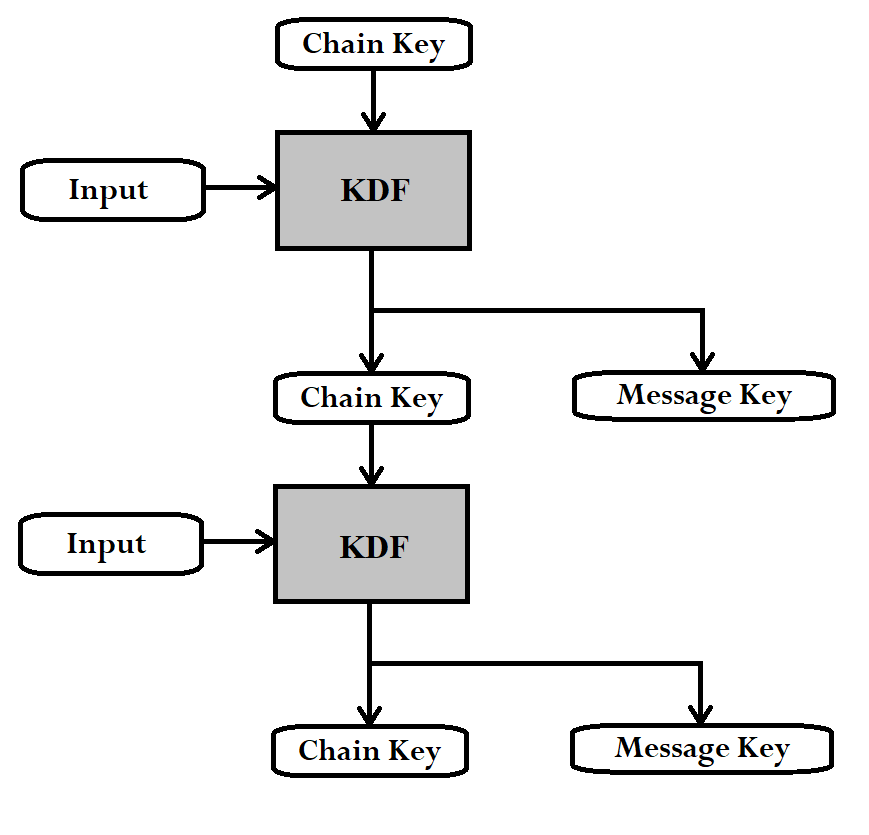}}
\caption{Symmetric Ratchet.}
\label{fig3}
\end{figure}

Fig. 3 shows the working principle of a symmetric key ratchet, at the heart of which lies the a KDF function that takes some input data and a random and secret KDF key (known as chain keys for the sending and receiving chains), and returns an output. A part of this output acts as a new KDF (chain) key and is used to replace the previous key and a remaining part of the key acts as a message key, which is used to encrypt the messages being exchanged using an Advanced Encryption Standard (AES) \cite{b7}. 

Now, the problem of using only a symmetric-key ratchet is that if the attacker (Mark) steals either Adam’s or Bud’s sending and receiving chain keys, he will be able to calculate all the future keys and hence decode all the future messages. In order to prevent this, the symmetric-key ratchet is combined with a DH-ratchet so as to update the chain keys depending upon the DH outputs. Fig. 4 explains the working principle of a Diffie-Hellman ratchet, where the DH function takes the user’s private key and the other party’s public key as the input and returns a DH output, which is used to derive a sending/receiving chain that matches exactly with the other party’s receiving/sending chain.  

\begin{figure}[htbp]
\centerline{\includegraphics[width=70mm]{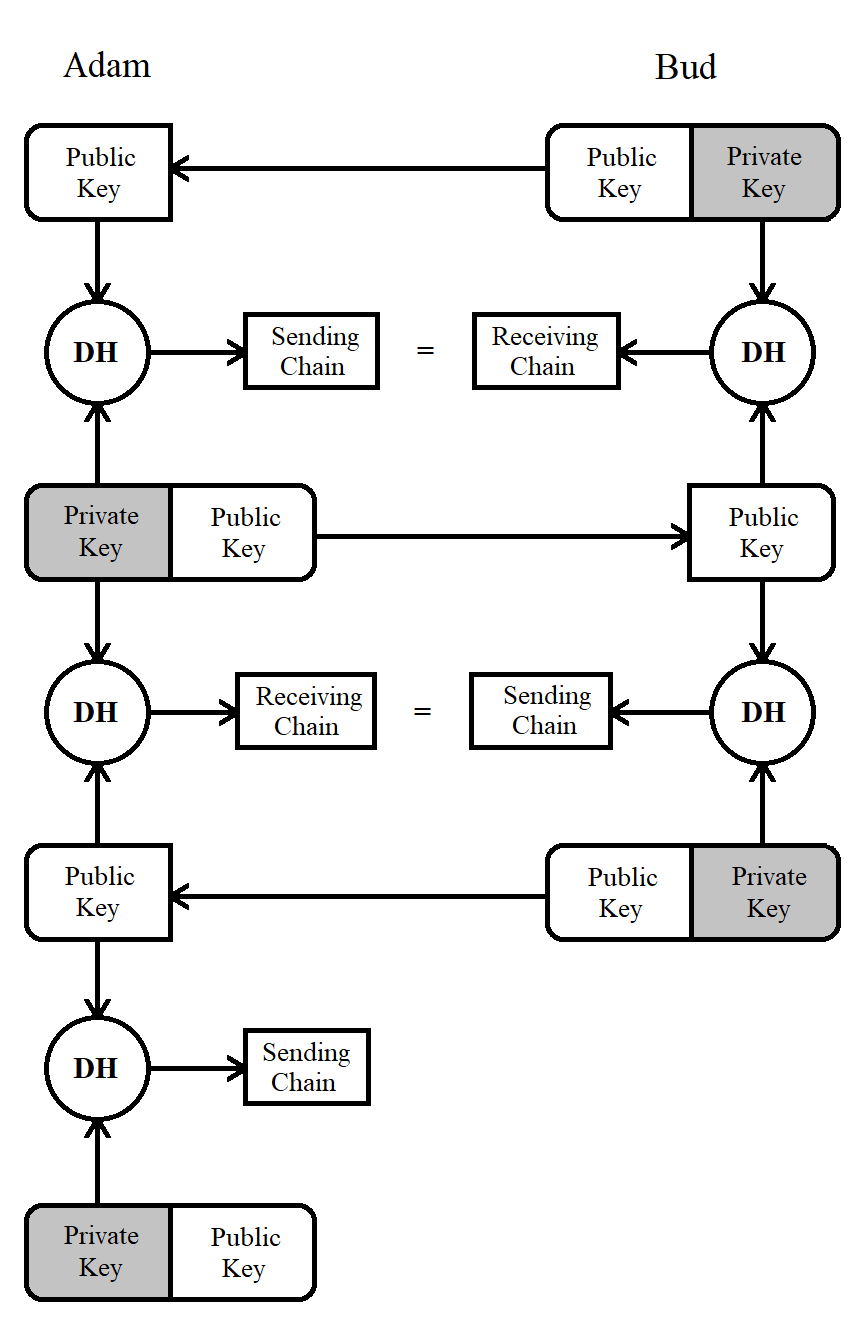}}
\caption{Diffie-Hellman Ratchet.}
\label{fig4}
\end{figure}

Whenever a new message is received or sent, the receiving or the sending chain is subjected to the symmetric-key ratchet step for obtaining the message key. And, whenever a user receives a new ratchet public key from the other party, a Diffie-Hellman ratchet step is executed before performing the symmetric-key ratchet in order to update the chain keys. A double ratchet is obtained by combining both the symmetric-key ratchet and the Diffie-Hellman ratchet together in a way such that the symmetric ratchet receives the output of the Diffie-Hellman ratchet as its input, thus producing a new Message key which is used for encrypting the message.

\section{Putting Everything Together}

Now, assuming that everything works properly and as expected, the following sequence of steps occur when Adam sends a message to Bud over WhatsApp. But, at first, both Adam and bud need to install WhatsApp in their phone, thus registering with the WhatsApp server. This includes signed prekey registration as well. 

\subsection{On Adam’s Side}
\begin{itemize}
\item Setting up the session using X3DH.
\item Computing the master secret and with help of the DH Ratchet step, finding out the root key as well as the chain key.
\item Deriving the next chain key (for the next step) and a message key with the help of the Hashing Ratchet.
\item Encrypting the message with the message key thus obtained, using AES256 in CBC mode.
\item The Hashing Ratchet moves forward by one step as Adam sends a message.
\item Whenever, he receives a message back from Bud, his DH Ratchet moves forward, thus computing a new chain key along with a new root key. 
\end{itemize}

\subsection{On Bud’s Side}
\begin{itemize}
\item Upon receiving the very first message from Adam, Bud completes setting up the session by computing the master secret and the root key, and from that the chain and the message keys. \item The message key thus obtained is used to decode the encoded message received from Adam.
\item Now, if Bud wants to sends a message back to Adam, he needs to generate a new pair of ephemeral keys, and then moves forward his DH Ratchet by one step with the help of the private key of the ephemeral pair and the root key, thus generating a new chain key and a message key.
\item This new message key thus obtained is used for encrypting the new message that Bud wants to send to Adam.
\item Bud sends this encrypted message to Adam with his ephemeral key in the header. 
\end{itemize}

\begin{figure}[htbp]
\centerline{\includegraphics[width=52mm]{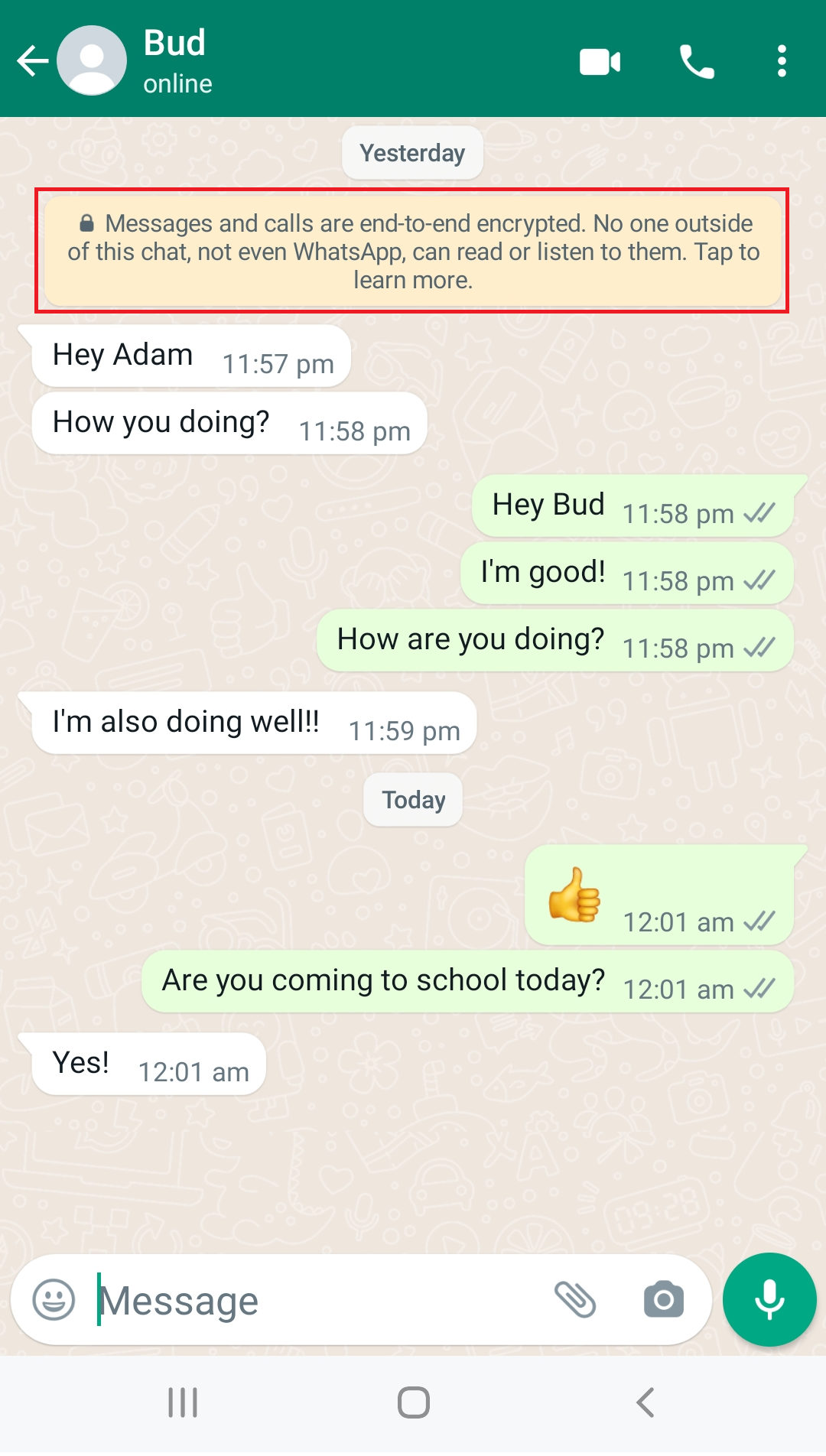}}
\caption{WhatsApp conversation of Adam with Bud.}
\label{fig5}
\end{figure}

\subsection{Example}
Fig. 5 shows an example of Adam’s WhatsApp conversation with Bud (as it appears on Adam’s phone), where the sequence of steps can be interpreted as follows:
\begin{enumerate}
\item \textbf{White$\,\to\,$White:} Bud’s Hashing Ratchet moves one step forward, thus generating the next chain key and a new message key. 

\item \textbf{Green$\,\to\,$Green:} Adam’s Hashing Ratchet moves one step forward, thus generating the next chain key and a new message key.

\item \textbf{White$\,\to\,$Green:} Adam’s DH Ratchet moves one step forward, thus generating the chain key and a new root key.

\item \textbf{Green$\,\to\,$White:} Bud’s DH Ratchet moves one step forward, thus generating the chain key and a new root key.
\end{enumerate}

\section{Privacy Concerns}
Though WhatsApp's End-to-End encryption system looks promising enough, and almost unbreakable, there are a few things we need to be concerned about – 
\vspace{1.5mm}
\begin{itemize}
\item If we believe that WhatsApp follows exactly the same specifications as Signal (which they claim they do) then there is no way to break the code, but unlike Signal, WhatsApp’s code is not public/open source, and thus there is no way to verify this. 
\vspace{1mm}
\item If we still assume that it follows exactly the same specifications as Signal, WhatsApp server can still find out who a particular user interacts with, how often, and how recently. 
\end{itemize}

\begin{figure}[htbp]
\centerline{\includegraphics[width=52mm]{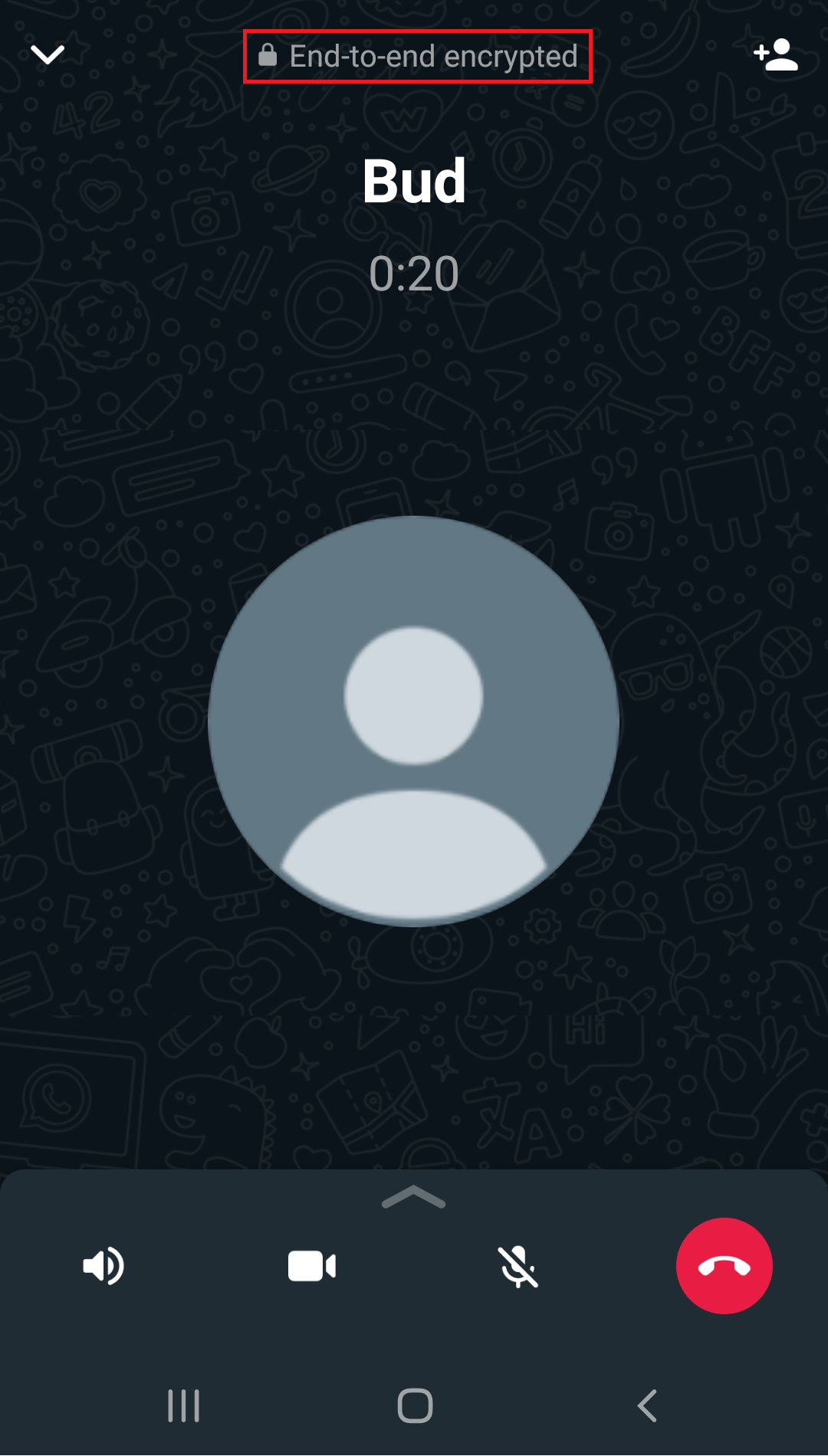}}
\caption{WhatsApp Call between Adam and Bud.}
\label{fig6}
\end{figure}

\section{Conclusion}
This paper describes the steps involved in the End-to-End Encryption system followed by WhatsApp sequentially, and explains each step and the corresponding terms in detail, along with mathematical expressions, examples and flowcharts as and when required. Though this paper focuses only on the text messages, WhatsApp also has the facilities of calling, video calling and exchanging media files – such as images, video, audio, documents, PDF etc. And all these facilities are End-to-End encrypted, e.g., Fig. 6 shows a screenshot of Adam's phone while being on a WhatsApp Call with Bud, which is End-to-End Encrypted. Besides describing the important and popular features of WhatsApp, this paper also highlights the privacy concerns associated with this messenger.

\vspace{0.5mm}

\section*{Acknowledgment}

I would like to thank Professor Dr. Grant Williams for beautifully explaining the basics of the cryptography in his lectures, which helped me understand the underlying concepts well and also for his constant help, guidance and feedback which helped me giving this paper a perfect shape.

\end{document}